# A crystal cleavage mechanism for UHV STM.

*A.I. Oreshkin, D.A. Muzychenko, I.V. Radchenko, V.N. Mantsevich, V.I. Panov and S.I. Oreshkin*[*]

119991 Moscow, Leninskie gory, Moscow State University (MSU), Physics Department, Russia
*119991, Moscow, Universitetsky pr., 13, Sternberg Astronomical Institute, MSU, Russia.

A device for UHV cleavage of crystal specimens for the use with STM has been suggested and developed. We present a device suitable for the precise cleavage of semiconductors. The device needs only small space and can be easily mounted in a small and compact UHV chamber equipped with a wobble stick manipulator. In order to prove the technique UHV STM measurements on *InAs(110)* surfaces with different bulk conductivities (p-and n-type) have been performed.

## 1. Introduction

Today scanning tunneling microscopy and spectroscopy (STM, STS) have become one of the traditional methods for surface investigations at atomic scale [1-5]. For the sample investigation under ultra high vacuum (UHV) conditions it is necessary to clean the surface in UHV systems. The following main methods are used to clean surfaces *in situ*:

1. Cleavage.
2. Thermal annealing.
3. Ion bombardment.
4. Chemical treatment.

A desorption of weak bonded particles from the surface could be realized by application of the thermal annealing only. However, in the case of strong bonded adsorbate particles, an application of repeated cycles of ion bombardments with a following annealing is usually necessary in order to clean surfaces. This method is the most useful especially for metallic surfaces. The chemical treatment method is suitable for the sample preparation in an environment of gases reacted with surface adsorbate under low pressure (approximately $10^{-6}$ torr). The formed weakly bonded molecules can be then thermally desorbed. The cleavage of single crystals *in situ* under UHV conditions or at low temperatures is the most favorable method to study the localized impurity states on semiconductor surface [6-8].

Two main types of cleavage principles exist in order to get flat surfaces. The first one is based on the press mechanism, and the second one on the shock mechanism. Worthy of mention is that for different samples different cleavage mechanisms have been constructed to obtain high quality surfaces. The cleavage device, based on press mechanism, is more effective for a large cleavage face preparation. Some additional manipulations for mutual orientation of the sample and cleavage tool should be performed *in situ* before cleaving. The shock mechanism in the cleavage device is suitable for a small cleavage face. Several different designs of the crystal cleavage mechanism are presented in [9-14]. The device for cleavage of magnesium oxide and alkali halide crystals of cross sections 5 mm×5 mm for use with LEED-Auger apparatus is described in [9]. The UHV single crystal cleavage apparatus for study NiO monocrystalline faces with AES, LEED and RHEED, adaptable to a flange of small inner diameter (38 mm) is reported in [10]. In Ref. [12] the cleaver is fixed on a flange (CF 35 UHV) that can be placed anywhere along the path of rotation-translation feedthrough for the sample transfer. The second manipulator is used to install the sample in the cleavage device only. The crystal cleavage device suggested in[14], consists of two manipulators. The device in Ref. [14] was used for the surface preparation of hard ionic crystals (MgO) with high precision[13].

Within the framework of UHV study of impurity atoms on $A_3B_5$ semiconductors surface with LEED and STM we constructed the new cleavage device based on the shock mechanism. The main distinctive feature of our device is its small size, which permits to integrate it almost anywhere in a small and compact chamber, and that it needs only a wobble stick for operation.

## 2. Mechanical system for sample cleavage.

The sample cleavage device was installed inside an Omicron STM UHV chamber. The following requirements for the sample cleavage have been satisfied with respect to the design of ourUHV system and sample transfer:

1. The mechanism of sample cleavage should be placed in the effective area of wobble-stick manipulator assembled in the STM chamber of UHV system. No additional manipulator is required to install the sample in the cleavage device.
2. The sample's fasteners should be performed on regular Omicron sample holders. Therefore it is easy to move the sample between STM, cleavage mechanism, rotation-translation feedthrough and carousel.

3. The sample cleavage mechanism should be suitable for many times cleaving before opening the UHV system. A sample in one sample holder can be cleaved only one time.

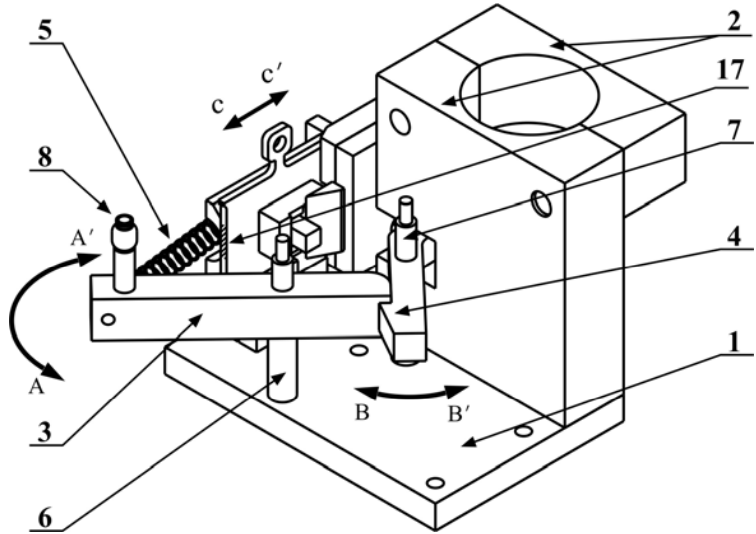

*Figure. 1. Main view of the cleavage system:*

*1-rigid plate, 2-fixing clamp, 3-hammer, 4-trigger, 5-spring, 6, 7-axes of hammer and trigger, 8-handle, 17-front edge of sample holder*

The general scheme of the cleavage mechanism with installed sample holder is presented on Fig.1. All parts of the system are assembled on the rigid plate 1 and attached by means of clamp 2 to one of the four rods on which STM suspension is arranged. The releasing mechanism consists of hammer 3, trigger 4 and spring 5. The hammer and the trigger can rotate freely around axes 6 and 7. These axes are fixed between the rigid plate 1 and the top cover (it is not indicated on Fig.1). The resetting of the releasing mechanism can be performed by means of the wobble-stick manipulator. The hammer is rotated by the manipulator in the *A'A* direction with help of handle 8 up to fixing by the trigger. The trigger is slightly spring-loaded in the *B* direction and moves in the *BB'* direction until it latches up. In order to cleave a sample, the trigger 4 is shifted in the *BB'* direction (Fig.1) by the wobble-stick manipulator. The hammer moves in the *AA'* direction under the spring force and strikes on the front edge 17 of the sample holder. The sample holder shifts in the *CC'* direction and a part of sample is cleaved. After cleavage the sample holder can shift no more than 0.1 mm in the *CC'* direction, which is to prevent the surface from damage by the edge of the knife. The tool angle of the knife was determined experimentally. The appropriate angle for $A_3B_5$ semiconductors cleavage is 50°- 60°. The design of the sample holder is based on the regular sample holder for the Omicron system, so it is easy to manipulate with a sample inside UHV system. Fig.2 displays the main scheme of the sample holder. The principle of the sample holder is not new and is already described in [14]. Sample 13 is attached to holder 12 by means of vice clamp 14. The vice clamp is attached to sample holder by plate 15 and screw. The sample is griped in the vice clamp by means of screw 16. Application of one screw helps us to avoid a sample stress during it's fixing in a vice. The position of the sample holder can be changed only in the horizontal direction in the plane normal to STM tip. This is why we installed the vice clamp on the sample holder horizontally to use thin plates of semiconductors crystals to prepare the samples. Moreover the position of the sample front surface (relative to the knife) edge is fixed during sample installation. Therefore a geometry size of cleaving crystals is of no importance and restricted only by the size of vice clamp and the spring force. Our cleavage device is completely surrounded by a tantalum basket in order to protect the STM from shattering samples.

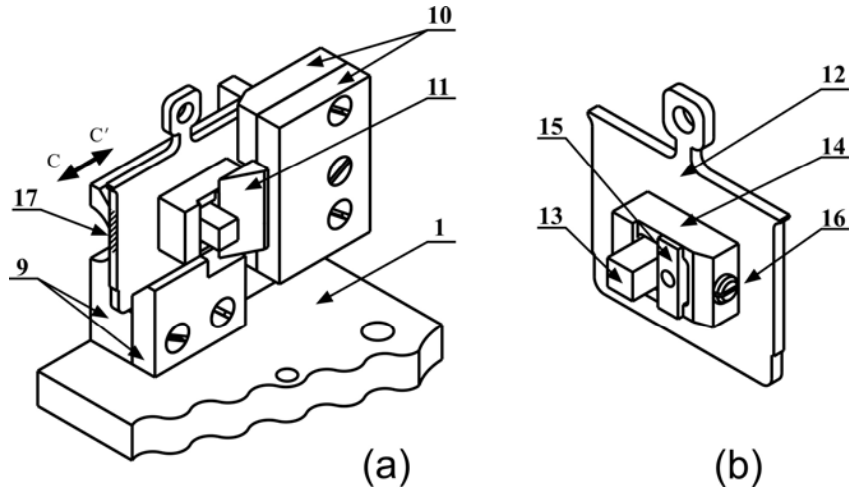

*Figure 2.* (a) - Mortise for the sample holder and the knife holder (b) - The sample holder.

*1-rigid plate, 9-mortise for sample holder, 10-knife holder, 11- knife, 12-sample holder, 13-sample, 14-sample clamp, 15-clamping plate, 16-clamping screw, 17-front edge of sample holder*

### III. Experimental results.

The surface of *InAs(110)* with n-type (p-type) bulk conductivity was studied *in situ* by STM. The surface of heavily doped $A_3B_5$ semiconductor single crystals with n-type bulk conductivity is the simplest case for STM imaging compared to other semiconductors surfaces [15]. On a relatively big scale (~200 Å) the surface is flat, and the individual doping atoms are clearly resolved. The *InAs* single crystals doped with *S* have been investigated. The chemical doping concentration was about $5 \times 10^{17}$ cm$^{-3}$. The *S* atoms on STM images have a round shape with a radius of approximately 30 Å.

They almost overlap resulting in the metallic conductivity of these samples even at liquid helium temperatures. The typical high resolution STM image of the *S* individual impurity on the *InAs(110)* surface is shown on Fig.3a. At negative sample bias voltage, the individual *S* impurity atom is imaged by STM as a round depression superimposed on the atomic lattice of the relaxed *InAs(110)* surface. A typical STM image of the *Zn* doped *InAs(110)* surface can be seen in Fig.3b. The bright spots can be identified as *Zn* impurity atoms. Their localization radius is about 20 Å, slightly less than that of the *S* impurity atom (30 Å) on the *InAs(110)* surface. In the vicinity of the impurity atoms the local density of state is strongly disturbed. Our experimental data are in a good accordance with results reported in [16].

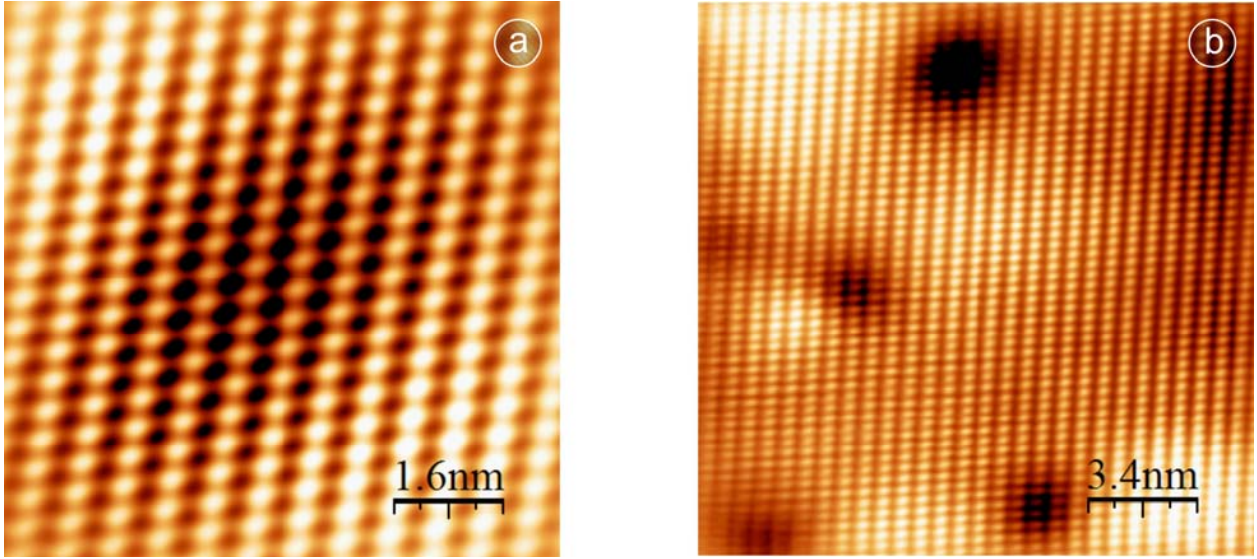

*Figure 3.* (a) Constant current STM image of S doped InAs(110), U=-1 V; I=53 pA.
(b) Constant current STM image of Zn doped InAs(110), U=-1.1 V; I=430 pA

### 4. Discussion.

In this paper a new design for a sample cleavage device has been described. The advantage of the design is that the device is small so that it can be integrated in any small UHV chamber. Further, only a wobble stick is needed for operation. STM results on the *InAs(110)* surface performed after cleavage have shown that our sample cleavage mechanism is suitable for the preparation of clean high quality semiconductor surfaces. Thus, it is possible to extract the information about a position, a concentration of the dopant atoms on the *InAs* cleavage surfaces from the STM images.

### 5. Acknowledgements.

This work was partially supported by RFBR grants № 06-02-17076-a, 06-02-17179-a, 05-02-19806-MF, president grant for scientific school No. 4599.2006.2. Support from Samsung corporation is also gratefully acknowledged. Authors are thankful to V.V. Gubernov, S.V. Savinov, A.A. Ezhov and N.S. Maslova for valuable discussions. We would like to express great thanks to the referee for his/her valuable remarks and corrections.